\documentclass[11pt]{article}
\usepackage{epsfig}

\author{A. Mondrag\'on and E. Hern\'andez\thanks{This work was
    partially supported by CONACYT (M\'exico) under contract 
    No. 4964-E9406}\\
  Instituto de F\'{\i}sica, UNAM\\
  Apartado Postal 20-364, 01000 M\'exico D.F. MEXICO}
\title{Accidental Degeneracy and Berry Phase of Resonant States}

\date{ 
}

\begin{document}

\maketitle
\begin{abstract}
  We study the complex geometric phase acquired by the resonant states
  of an open quantum system which evolves irreversibly in a slowly
  time dependent environment. In analogy with the case of bound
  states, the Berry phase factors of resonant states are holonomy
  group elements of a complex line bundle with structure group $\bf
  C^*$. In sharp contrast with bound states, accidental degeneracies
  of resonances produce a continuous closed line of singularities
  formally equivalent to a continuous distribution of ``magnetic''
  charge on a ``diabolical'' circle, in consequence, we find different
  classes of topologically inequivalent non-trivial closed paths in
  parameter space.

\end{abstract}

\section{Introduction.}

For many years now, it has been appreciated that there are distinct
advantages in describing quantum resonances and the quantum phenomena
associated to the production, evolution and decay of resonances in
terms of resonant or Gamow states, since many physical effects are
then readily expressed and evaluated\cite{one}. In this work, we will
give closed analytical expressions for the complex Berry phase of an
open quantum system in a resonant state of a Hermitian Hamiltonian
with non-self-adjoint boundary conditions, and we will discuss some of
its properties.

During the last fourteen years the geometric phase factors arising in
the adiabatic evolution of quantum systems\cite{two} have been the
subject of many investigations\cite{three,four}. The early literature
was mostly concerned with the geometric phase factors of closed
systems driven by Hermitian Hamiltonians\cite{four}. More recently
there has been a substantial interest in the complex geometric phase
acquired by the eigenstates of open quantum systems. This problem
arises naturally in connection with various experiments which, by
their very essence, require the observation of the geometric phase in
metastable states. The Berry phase in the optical supermode
propagation in a free laser, which is a classical system described by
a Schr\"odinger-like equation with a non-Hermitian Hamiltonian, was
studied by Dattoli et al.\cite{five}. The measurement of the geometric
phase in atomic systems with two energy levels, one of which at least
is metastable, was also described in terms of a non-Hermitian
Hamiltonian by Miniatura et al.\cite{six}. The validity of the
adiabatic approximation for dissipative, two level systems driven by
non-Hermitian Hamiltonians was examined by Nenciu and
Rasche\cite{seven}, and by Kvitsinsky and Putterman\cite{eight}, who
also established that the Berry phase is complex in this case.
Sun\cite{nine} proposed a higher-order adiabatic approximation for
two-level non-Hermitian Hamiltonians, and showed that the holonomy
structure associated to the Berry phase factor in this non-Hermitian
case is the non-unitary generalization of the holonomy structure of
the Hermitian case.

In two previous papers\cite{ten,eleven}, we gave explicit expressions
for the geometric phase of true resonant states, defined as complex
energy eigenstates of a Hermitian Hamiltonian which satisfy purely
outgoing wave boundary conditions at infinity\cite{twelve}, and
pointed out some of the mathematically interesting and physically
relevant properties resulting from the extended nature of the
singularities in parameter space associated with the occurrence of
accidental degeneracies of two resonances. In this conection, in
another paper we showed that the codimension of the accidental
degeneracy of $n$ resonances differ significantly from those of bound
states, we also showed that, close to a crossing of two resonant
states, the topological structure of the energy hypercomplex surfaces
differ significantly from the double conical point singularity typical
of bound states\cite{thirteen}. Later, by means of a numerical
analysis of the experimental data on the 2$^+$ doublet of resonances
with $T = 0,1$ in the energy spectrum of $^8$Be, we showed, in a
realistic example, that a true crossing of resonances mixed by a
Hermitian interaction may be brought about by the variation of only
two real linearly independent parameters\cite{fourteen}. In this paper
we show that the geometric interpretation of the Berry phase factor,
first given by B. Simon\cite{fifteen} for the adiabatic evolution of
closed quantum systems, may be generalized, in a very natural way, to
the case of resonant states of open quantum systems. That is, the
adiabatic evolution of resonant states may be interpreted as parallel
transport in a complex line bundle defined over the space of
parameters with structure group $\bf C^*$ (the multiplicative group of
the non-zero complex numbers). Then, the Berry phase factors of
resonant states arise as the holonomy group elements due to a
connection in the bundle such that during the adiabatic evolution the
resonant state is parallel transported along the fibre. The horizontal
spaces are perpendicular to the fibre with respect to a generalized
inner product of resonant states defined in a rigged Hilbert space.

\section{Resonant States}
\subsection{Resonant states in a slowly time evolving environment}

Let us consider the time evolution of a quantum system in a state
which is a superposition of unstable eigenstates moving in some strong
external field of force which changes slowly with time. In order to
have some concrete example in mind, although a very hypothetical one,
we may think of an $^8Be$ nucleus which has only unstable energy
eigenstates moving in the field of forces of a double magic nucleus,
like $^{208}Pb,$ in a peripheral collision in which the distance
between the two nuclei is never smaller than the sum of the nuclear
radii. In a semiclassical treatment of the collision, when the centers
of the nuclei move along classical trajectories, the parameters in the
nucleus-nucleus interaction change with time\cite{sixteen}.

The evolution of the system under the influence of the external
perturbation is governed by the time dependent Schr\"odinger equation

\begin{equation}
\label{uno}i\hbar \frac{\partial \Psi }{\partial t}=H\Psi . 
\end{equation}
The Hamiltonian $H$ is the sum of the time-independent Hamiltonian
$H_0$ describing the evolution of the unperturbed system plus a
perturbation term $H_1$ which is a function of a number $N$ of
``external" parameters $\left\{ X_1,X_2...X_N\right\} $ which may
change with time,

\begin{equation}
\label{dos}H(t) = H_0 + H_1(X_i(t)).
\end{equation}

The energy eigenfunctions of the unperturbed Hamiltonian are the
solutions of the equation,

\begin{equation}
\label{tres}H_0\varphi _m(\xi _i,\eta _j, r)={\cal E}_m
\varphi _m(\xi _i,\eta _j, r),
\end{equation}

\noindent
the wave functions $\varphi _m(\xi _i,\eta _j, r)$ satisfy the
boundary conditions appropriate to a decaying state.

Assuming that, in the absence of perturbation, the unstable system
decays spontaneously in two stable fragments, the unperturbed energy
eigenfunctions may be written as cluster model wave
functions\cite{seventeen}

\begin{equation}
\label{cuatro}\varphi _m(\xi _i,\eta _j, r)={\cal A}
\left\{ \phi _A(\xi _i)\phi _B(\eta _j)\frac{u_{ml}
(r)}{r}{\cal Y}_{Jls}^M(\hat r)\right\} . 
\end{equation}

\noindent\
where $\phi _A(\xi _i)$ and $\phi _B(\eta _j)$ are the wave functions
of the clusters A and B, $u_{ml}(r)$ is the radial part of the wave
function of the relative motion of the two clusters, ${\cal
  Y}_{Jls}^M(\hat r)$ is a spherical harmonic and ${\cal A}$ is the
antisymmetrizer. In our example, $\varphi _m$ would be the
eigenfunction of a state of $^8$Be which decays spontaneously in two
$^4$He clusters. In this case $u_{ml}(r)$ is a Gamow function.

\subsection{A few facts about Gamow functions}

Gamow functions are the eigenfunctions of the time independent
Schr\"odinger equation which vanish at the origen, and behave as
purely outgoing waves for large values of the relative distance $r$,

\begin{equation}
\label{cinco}u_{ml}(0) = 0,
\end{equation}

\noindent\
and

\begin{equation}
\label{seis}\lim_{r\rightarrow\infty} \left[ u_{ml}(r) 
- O_l(k_m,r)\right] = 0, 
\end{equation}

\noindent\
where the function $O_l(k_m,r)$ is an outgoing spherical wave of
complex wave number $k_m$ and angular momentum $l$\cite{twelve}.

The boundary condition (\ref{seis}) is not self-adjoint, in
consequence, the energy eigenvalues are complex, with $Re{\cal E}_m>0$
and $Im{\cal E}_m< 0.$ Hence, Gamow functions are a generalization of
bound state eigenfunctions in that they belong to complex wave
numbers, $k_m = \kappa _m - i\gamma _m$, with $\kappa _m > \gamma _m >
0$, instead of purely imaginary ones. This generalization leads out of
the Hilbert space based quantum mechanics. Therefore, the quantum
mechanical rules for normalization, orthogonality and completeness in
their usual form, do not apply. Nevertheless, bound and resonant
states form a bi-orthonormal set with their adjoints, which may be
extended by a continuum of suitably chosen scattering states of
complex wave number to form a complete set in terms of which any
square integrable function may be expanded \cite{twelve}.

The symmetry properties of the Schr\"odinger equation and the boundary
conditions under the operations of complex conjugation and reversal of
time suggest a definition for the adjoint $\tilde u_{ml}(k_m,r)$ of
the Gamow function $u_{ml}(k_m,r)$\cite{twelve},

\begin{equation}
\label{siete}\tilde u_{ml}(k_m,r) = u^*_{ml}(-k^*_m,r).
\end{equation}

\noindent\
Now, if the adjoint $\tilde u_{ml}(k_m,r)$ is identified with the
bra-eigenfunction, the quantum mechanical inner product, or bra-c-ket
rule, may be generalized to include Gamow eigenfunctions. With this
prescription, matrix elements of operators and inner products are
computed in configuration space representation as integrals over the
radial variable $r$. Since the Gamow function $u_{ml}(k_m,r)$ and its
adjoint $\tilde u_{ml}(k_m,r)$ oscillate between envelopes that grow
exponentially with $r$, the integrals over $r$ must be properly
defined. This may be done, either by analytic continuation in the
complex $k$-plane from above \cite{twelve,eighteen} or by means of an
Abel regulator\cite{nineteen} with a suitable convergence factor and a
limiting procedure\cite{twenty,twentyone}. Both procedures give the
same result\cite{twentytwo,twentythree}. In this paper, we will adopt
the second definition.

Then, it may be shown that

\begin{equation}
\label{ocho}\lim_{\mu\rightarrow 0} \int_{0}^{\infty} e^{-\mu
  r^2}u^{*}_{ml}(-k^*_m,r)u_{nl'}(k_n,r)dr = \delta_{mn}\delta_{ll'}
\end{equation}

\noindent\
i.e, the Gamow eigenfunctions are orthonormalized in a generalized
sense. It may also be shown\cite{twelve,twentyfour} that, for any two
square integrable functions, $f(r)$ and $g(r)$, the following relation
holds

\begin{equation}
\label{nueve}
\begin{array}{c}
\int_{0}^{\infty} f^*(r)g(r)dr = \sum_n\left\{\left[
\int_{0}^{\infty} f^*(r)u_{nl}(k_n,r)dr\right]\times 
\left[\int_{0}^{\infty} u^*_{nl}(-k^*_n,r')g(r')dr'
\right]\right\} +\\ \int_{c} dk\left[\int^{\infty}_{0}
f^*(r)\phi_l(k,r)dr\right]\left[\int_{0}^{\infty}
\phi^*_l(-k^*,r')g(r')dr' \right].
\end{array}
\end{equation}

\noindent\
In this expression, the functions $u_{nl}(r)$ are bound states or
Gamow state eigenfuctions belonging to real negative or complex
eigenvalues ${\cal E}_n$, the functions $\phi_l(k,r)$ are scattering
partial wave functions of angular momentum $l$ and complex wave number
$k$. The integration contour $C$, in the wave number plane $k$, starts
from the origin as a straight line with slope -1, it goes down to $Im
k = \alpha $ and then, it continues as a straight line parallel to the
real axis\cite{twelve}. Finally, the square brackets around the
integrals mean that, when necessary, the integrals are defined by
means of a gaussian regulator and a limit as in (\ref{ocho}) or by
analytical continuation from above as in Mondrag\'on and
Hern\'andez\cite{twelve,twentyfour}.

Since $f(r)$ and $g(r)$ are arbitrary functions, we are justified in
writing the expansion

\begin{equation}
\label{diez}g(r) = \sum_{m} u_{ml}(r)<u_{ml}|g> + \int_c
\phi_l(k,r)<\phi_l(k)|g>dk,
\end{equation}

\noindent\
the index $m$ runs over bound and resonant states.

The expansion coefficients are given by

\begin{equation}
\label{once}<u_{ml}|g> = \lim_{\mu \rightarrow
  0}\int_{0}^{\infty}e^{-\mu r^2}u^*(-k^*_m,r)g(r)dr,
\end{equation}

\begin{equation}
\label{doce}<\phi_l(k)|g> = \lim_{\mu \rightarrow
  0}\int_{0}^{\infty}e^{-\mu r^2}\phi^*_l(-k^*,r)g(r)dr.
\end{equation}

Once the validity of the quantum mechanical inner product has been
generalized to apply to bound and resonant eigenfunctions of the
relative motion of the two clusters, the generalization of
(\ref{nueve}) and (\ref{diez}) to expansions of many body wave
functions in terms of cluster model bound and resonant eigenfunctions
is fairly straightforward\cite{twelve,twentyfour}.

\subsection{The mixing matrix}

We may, now, go back to our problem, namely the time evolution of the
resonant states of a many body system moving in a slowly
time-dependent external field of force. Since we are interested in the
time evolution of a state $\Psi$ which is a superposition of unstable
states, we make an expansion of the wave function $\Psi $ in terms of
bound and resonant states of $H_0,$

\begin{equation}
\label{trece}\Psi =\sum_ma_m(t)\varphi _m(\xi _i,\eta_j,r)+
\int_cb(k;t)\varphi^{(+)}(k;\xi_i,\eta_j,r )dk. 
\end{equation}

In general, the index $m$ runs over bound and resonant states. In our
example, $ \varphi _m$ would be the complex energy eigenfunctions of
the $^8$Be nucleus which decays spontaneously in two $^4$He clusters.
The scattering states $\varphi ^{(+)}(k,\xi _i,\eta_j,r)$ of complex
wave number $k$ and the integration contour $C$ are defined in the
previous subsection.

Substitution of (\ref{trece}) in (\ref{uno}) gives the set of coupled
equations

\begin{equation}
\label{catorce} 
\frac{d a_m(t)}{dt} = -\frac i\hbar {\cal E}_ma_m(t)-\frac i\hbar
\sum_n<\varphi _m|H_1(t)|\varphi _n>a_n(t)  -\frac i\hbar \int_c<\varphi
_m|H_1(t)|\varphi ^{(+)}(k)>b(k;t)dk 
\end{equation}
and a similar expression for $d b(k,t)/dt.$

We will use the notation $|\varphi _m(\xi _i,\eta_j,r)>$ for the Gamow
function, and $<\varphi _m(\xi _i,\eta_j,r)|$ for its adjoint. Hence,
the matrix element of the perturbation term $H_1(t)$ taken between
bound or resonant states of the unperturbed system is given by

\begin{equation}
\label{quince}\left\langle \varphi _m\left| H_1(t)\right| \varphi
_n\right\rangle =\int \cdot \cdot \cdot \int \langle \varphi _m(\xi
_i,\eta_j,r)|H_1(t)|\varphi _n(\xi_i,\eta_j,r)\rangle d^{3}\xi _1
\cdot \cdot \cdot d^{3}\xi _{A-1}d^{3}\eta_{1}\cdot \cdot \cdot 
d^{3}\eta_{B-1}d^3r, 
\end{equation}

\noindent\
the integral over the radial variable $r$ is defined as in
(\ref{ocho}).

When the interactions are time reversal invariant, the dual of the
complex Gamow function $u_{ml}(r)$ is the same
function\cite{twelve,twentyone,twentyfour}. But, when the interactions
are not time reversal invariant, the Gamow function and its dual are
not the same function.

Therefore, when the forces acting on the system are time reversal
invariant, the complex matrix $\bf H$, with matrix elements

\begin{equation}
\label{dieciseis}{\bf H}_{mn}(t) = {\cal E}_m\delta _{mn}+
\left\langle \varphi_m\left| H_1(t)\right| \varphi _n\right\rangle 
\end{equation}

\noindent
is symmetric but non-Hermitian. When the forces acting on the system
are not time reversal invariant, $\bf H$ is, in general, complex,
non-symmetric and non-Hermitian.

The contribution of the non-resonant background integral over the
continuum of scattering wave functions to (\ref{trece}) and
(\ref{catorce}) will not be relevant to the following discussion.
Therefore, to ease the notation, we will disregard the contribution of
the background of scattering functions. With this truncation,

\begin{equation}
\label{diecisiete}H\Psi =\sum_m|\varphi_m(\xi_i,\eta_j,r )> 
\left[ \sum_n{\bf H}_{mn}a_n(t)\right]. 
\end{equation}

Then, the set of coupled equations(\ref{catorce}) reduces to

\begin{equation}
\label{dieciocho}\dot a_m(t)=-\frac i\hbar \sum_n{\bf H}_{mn}a_n(t), 
\end{equation}
where $\dot a_m(t)$ is the time derivative of $a_m(t).$

\section{Geometric Phase of a Resonant State.}

The time evolution of the quantum system is governed by the
Hamiltonian $H$, or the matrix $\bf H$, which are functions of $N$
real, linearly independent parameters. Therefore, we may consider the
matrix $\bf H$ embedded in a population of $\bf H$ matrices smoothly
parametrized by $N$ external parameters which take values in some
domain $D$ of a manifold or parameter space. Each point in $D$
represents an $\bf H$ matrix. When the external parameters change with
time, the system traces a path $C$ in parameter space. In the
following, we will study the behaviour of the system in the time
interval [0,T], assuming that at the initial time, $t=0$, it is in an
eigenstate of $\bf H$, at $t=T$ the parameters $(X_1,X_2,....X_N)$
have returned to their initial values, and the adiabatic theorem
holds. Then, the unstable system traces a closed path $C$ in parameter
space while it remains in an eigenstate of ${\bf H}(t)$ at all times.

In the absence of symmetry, $\bf H$ has no repeated eigenvalues at
almost all points in the domain $D$. The set of points in $D$ where
$\bf H$ has one twofold repeated eigenvalue is a subdomain $D'\subset
D$. If $D$ has $N$ dimensions, and $D'$ has $N'$ dimensions, then
$N=N'+\kappa$, $\kappa$ is the codimension of the twofold degeneracy.
Similar relations hold in the case of an m-fold degeneracy.

It will be assumed that the complex non-Hermitian matrix $\bf H$ has
no repeated eigenvalues at all points on the path $C$. In consequence,
at all points on $C$, it may be brought to diagonal form by means of a
similarity transformation

\begin{equation}
\label{diecinueve}{\bf K}^{-1}{\bf HK}={\bf E}, 
\end{equation}

\noindent
where ${\bf E}$ is the diagonal matrix of the complex energy
eigenvalues. The columns in the matrix ${\bf K}$ are the instantaneous
right eigenvectors of ${\bf H}$. In an obvious notation

\begin{equation}
\label{veinte}{\bf K}=\left( |\phi ^{(1)}\rangle ,|\phi ^{(2)}\rangle
,......|\phi ^{(s)}\rangle ....|\phi ^{(n)}\rangle \right) . 
\end{equation}

\noindent
which satisfy the eigenvalue equation

\begin{equation}
\label{veintiuno}{\bf H}\left( X_l(t)\right) |\phi ^{(s)}(t)\rangle =
{\cal \hat E}_s(t)|\phi ^{(s)}(t)\rangle. 
\end{equation}

The rows in ${\bf K}^{-1}$ are the corresponding left eigenvectors of
${\bf H}$, properly normalized,

\begin{equation}
\label{veintidos}\left\langle \phi ^{(i)}|\phi ^{(j)}\right\rangle
=\delta _{ij}. 
\end{equation}

The adiabatic basis, $\left\{ \left|\hat \varphi _s\left(
      \xi_i,\eta_j,r;X_l(t)\right) \right\rangle \right\} ,$ of
instantaneous bound and resonant energy eigenstates of the complete
Hamiltonian $H$, is obtained from the set of unperturbed bound and
resonant states with the help of the matrix $\bf K$

\begin{equation}
\label{veintitres}|\hat \varphi _s(\xi _i,\eta_j,r;X_l(t))\rangle
=\sum_m|\varphi _m(\xi _i,\eta_j,r)\rangle {\bf K}_{ms}(t) 
\end{equation}
and their adjoints (duals) are given by

\begin{equation}
\label{veinticuatro}\langle \hat \varphi _s(\xi _i,\eta_j,
r;X_l(t)) |=\sum_n({\bf K}^{-1}(t))_{sn}\langle \varphi _n
(\xi _i,\eta_j,r)|. 
\end{equation}

It follows that the instantaneous bound and resonant energy
eigenstates of $H(t)$ satisfy an orthogonality relation similar to
(\ref{ocho}), which they inherit from the unperturbed complex energy
eigenfunctions

\begin{equation}
\label{veinticinco}
\begin{array}{c}
\langle \hat \varphi _s(\xi _i,\eta_j,r;X_l(t)) |\hat \varphi _{s'}
(\xi_i,\eta_j,r;X_l(t))\rangle = \sum_{m,n}({\bf K}^{-1})_{sn}
\langle \varphi_n(\xi_i,\eta_j,r)|\varphi_m(\xi_i,\eta_j,r)
\rangle {\bf K}_{ms'} = \\ \sum_{m,n}({\bf K}^{-1})_{sn}
\delta_{nm}{\bf K}_{ms'} = \delta_{ss'}.
\end{array}
\end{equation}

If we multiply both sides of eq.(\ref{veintitres}) by ${\bf K}^{-1}$,
we obtain

\begin{equation}
\label{veintiseis}|\varphi_m(\xi_i,\eta_j,r)\rangle = \sum_s|\hat 
\varphi_s(\xi_i,\eta_j,r;X_l(t))\rangle ({\bf K}^{-1})_{sm}.
\end{equation}

Substitution of this expresion in (\ref{trece}), gives the expansion
of $\Psi$ in instantaneous energy eigenfunctions of $\bf H$,

\begin{equation}
\label{veintisiete}\Psi =\sum_s|\hat \varphi _s\left( \xi_i,
\eta_j,r;X_l(t)\right) \rangle \hat a_s(t), 
\end{equation}
where

\begin{equation}
\label{veintiocho}\hat a_s(t)=\sum_n\left( {\bf K}^{-1}(t)\right)
_{sn}a_n(t). 
\end{equation}

\noindent\
In (\ref{veintisiete}) we have kept only the summation over bound and
resonant states and, as in the previous section, we have disregarded
the contribution from the integral over the continuum of scattering
functions of complex wave number.

Similarly, the expansion of $H\Psi $ becomes

\begin{equation}
\label{veintinueve}H\Psi =\sum_s|\hat \varphi _s\left( \xi_i,\eta_j,
r;X_l(t)\right) \rangle {\cal \hat E}_s(t)\hat a_s(t). 
\end{equation}

Substitution of (\ref{veintisiete}) and (\ref{veintinueve}) in the
time dependent Schr\"odinger equation gives the set of coupled
equations

\begin{equation}
\label{treinta}\frac{d\hat a_s(t)}{dt}+\sum_{m=1}\langle \hat
\varphi _s|\nabla _\Re\hat \varphi _m\rangle \cdot \frac{d\vec \Re}
{dt}\hat a_m(t)=-i{\cal \hat E}_s(t)\hat a_s(t) 
\end{equation}

\noindent\
where, $|\nabla_\Re\hat\varphi_m >$ is the gradient of $|\hat\varphi_m
>$ in parameter space,

\begin{equation}
\label{treintaiuno}\frac{d|\hat\varphi_m >}{dt} = \sum_{i=1}^N
\left(\frac{\partial}{\partial X_i}|\hat\varphi_m >\right)
\frac{dX_i}{dt} = |\nabla_{\vec \Re}\varphi_m >\cdot 
\frac{d \vec \Re}{dt}
\end{equation}

It will be assumed that the non-adiabatic transition amplitudes are
very small

\begin{equation}
\label{treintaidos}\frac 1{\left| \hat a_s\right| }\left| 
\langle \hat \varphi _s|\nabla _\Re\hat \varphi _m\rangle 
\cdot \frac{d\vec \Re}{dt}\right| <<1,\qquad m\neq s. 
\end{equation}

Then, we can make the approximation

\begin{equation}
\label{treintaitres}\frac 1{\hat a_s}\frac{d\hat a_s}{dt}
\simeq -i{\cal \hat E }_s(t)-\langle \hat \varphi _s|
\nabla _\Re\hat \varphi _s\rangle \cdot \frac{d\vec \Re}{dt}. 
\end{equation}
Integrating both sides, we get

\begin{equation}
\label{treintaicuatro}\hat a_s(t)=exp [ {-\frac i\hbar 
\int_{t_0}^t{\cal \hat E}_s(t^{\prime })dt^{\prime }} ] 
exp [ {i\gamma _s}\hat] a_s(0), 
\end{equation}
the first factor is the complex dynamical phase, whereas the second
one is the complex Berry phase given by

\begin{equation}
\label{treintaicinco}\gamma _s=i\int_{{\bf c}}\langle 
\hat \varphi _s|\nabla _\Re\hat \varphi _s\rangle \cdot d\vec \Re. 
\end{equation}

Direct evaluation of $\nabla _\Re|\hat \varphi _s\rangle$ requires a
locally single valued basis for $|\hat \varphi _s\rangle$ and can be
awkward. Such difficulties are avoided by transforming the path
integral into a surface integral with the help of Stokes theorem

\begin{equation}
\label{treintaiseis}\gamma _s=i\sum_{m\neq s}\int_\Sigma 
\int_{\partial \Sigma ={\bf c}}\langle \nabla_\Re \hat \varphi _s
|\hat \varphi _m\rangle \wedge \langle \hat \varphi _m|\nabla _\Re
\hat \varphi _s\rangle \cdot d\vec \Sigma, 
\end{equation}
where $\Sigma $ is any surface in parameter space whose boundary is
the curve ${\bf C}$, and $\wedge$ means wedge product.

This expression may be written in a more convenient form by means of 
the identity

\begin{equation}
\label{treintaisiete}\langle \hat \varphi _s|\nabla _\Re
\hat \varphi _m\rangle =\frac 1{{\cal \hat E}_m-{\cal 
\hat E}_s}\langle \hat \varphi _s|\nabla_\Re H_1|
\varphi _m\rangle , 
\end{equation}
then,

\begin{equation}
\label{treintaiocho}\gamma _s=i\sum_{m\neq s}\int_\Sigma 
\int_{\partial \Sigma ={\bf c}}\frac{\langle \hat \varphi _s|
\nabla _\Re H_1|\hat \varphi _m\rangle \wedge \langle 
\hat \varphi _m|\nabla _\Re H_1|\hat \varphi _s\rangle
\cdot d\vec \Sigma }{\left( {\cal \hat E}_s-{\cal \hat E
}_m\right) ^2}, 
\end{equation}
provided the surface $\sum $ does not cross any point in $D$ where the
denominator vanishes. Since the dependence on $|\nabla _\Re \varphi
_s\rangle$ has been eliminated in (\ref{treintaiocho}), phase
relations between eigenstates with different parameters are no longer
important, any complete set of instantaneous eigenfunctions of the
time dependent Hamiltonian $H$ may be used to evaluate the integral in
(\ref{treintaiocho}).

Explicit expressions for the Berry phase in terms of our
parametrization of the interaction Hamiltonian may easily be obtained.
Since the unperturbed Hamiltonian and its bound and resonant states
are independent of time, the time dependence of the bound and resonant
instantaneous energy eigenstates of the perturbed system is entirely
contained in the matrix $\bf K$, which is a function of $t$ through
the time dependence of the external parameters.  Therefore, from
eqs.(\ref{veintitres}), (\ref{veinticuatro}) and (\ref{veinticinco})
we get

\begin{equation}
\label{treintainueve}\gamma _s=i\int_{{\bf c}}\left[ {\bf K}^{-1}
(\nabla _\Re{\bf K})\right] _{ss}\cdot d\vec \Re. 
\end{equation}

The columns in $\nabla _\Re \bf K$ are the gradients of the
instantaneous eigenvectors of ${\bf H}(t)$. Therefore, $\gamma _s$ may
also be written as

\begin{equation}
\label{cuarenta}\gamma _s=i\int_{{\bf c}}\left\langle \phi _s
|\nabla_\Re\phi _s\right\rangle \cdot d\vec\Re. 
\end{equation}

Furthermore, from (\ref{treintaiseis}) we may derive a sum rule for
the geometric phases of the interfering resonant states. Writing the
integrand in the right hand side of (\ref{treintaiseis}) in terms of
$\bf K$ and $\nabla_\Re{\bf K}$, and taking the sum over $s$, we get

\begin{equation}
\label{cuarentaiuno}\sum_s\gamma_s = i\int_{\Sigma}\int_{\partial 
\Sigma=c}tr\left[(\nabla_\Re{\bf K}^{-1})\wedge(\nabla_\Re{\bf K})
\right]\cdot d\Sigma
\end{equation}

\noindent\
but

\begin{equation}
\label{cuarentaidos}-\frac{1}{2\pi}tr\left[(\nabla _\Re\bf K^{-1}
)\wedge (\nabla_\Re \bf K)\right] = C_1(N),
\end{equation}

\noindent\
is the first Chern class of a complex vector bundle defined by the
matrix $\bf H$ over the parameter space M. Its integral is the first
Chern number\cite{twentyfive}

\begin{equation}
\label{cuarentaitres}-\frac{1}{2\pi}\int_{\Sigma}\int_{\partial 
\Sigma = c}tr\left[(\nabla_\Re \bf K^{-1})\wedge(\nabla_\Re \bf K)
\right ]\cdot d\vec \Sigma  = c_1.
\end{equation}
\noindent
Hence,

\begin{equation}
\label{cuarentaicuatro}\sum_s\gamma _s=i\oint tr\left[ 
{\bf K}^{-1}\left( \nabla _\Re{\bf K}\right) \right] \cdot d
\vec \Re. =-2\pi c_1,
\end{equation}

\noindent
is a topological invariant.

\section{Berry phase factors of resonant states and holonomy 
in a complex line bundle}
\subsection{Resonant states as elements of a rigged Hilbert space}

Since 1983, B. Simon\cite{fifteen} pointed out that for Hermitian
closed quantum systems, the adiabatic evolution can be interpreted as
a parallel translation in a Hermitian line bundle and the Berry phase
factor is the holonomy in such a bundle. In this section , it will be
shown that this geometric interpretation may be generalized in a very
natural way to the Berry phase factor of resonant states of open
quantum systems.

The resonant or Gamow functions $\left\{|\phi _m(\xi_i,\eta_j,r
  >\right\}$ of the unperturbed system are eigenfunctions of the
self-adjoint unperturbed Hamiltonian $H_0$, which satisfy purely
outgoing wave boundary conditions for large values of the separation
distance $r$ of the decay fragments. Since the boundary condition is
not self-adjoint, the corresponding energy eigenvalues ${\cal E}_m$,
are complex, with $Re {\cal E}_m>$0 and $Im {\cal E}_m<$0. In
configuration space representation, the Gamow functions, as functions
of $r$, behave as outgoing waves which oscillate between envelopes
that increase exponentially. In consequence, Gamow functions are not
square integrable and cannot be characterized as elements of a Hilbert
space.

Gamow functions are usually associated with the resonance poles of the
scattering matrix and the resolvent operator of the time independent
Schr\"odinger equation which lie in the lower half-plane of the
unphysical sheet of the Riemann energy surface\cite{one,twelve}. In
order to give a proper mathematical characterization of Gamow
functions as elements of a space, one has to specify a rigged Hilbert
space in which the Gamow states $\left\{|\varphi_m(\xi_i,\eta_j,r)
  >\right\}$ are defined as generalized eigenvectors of the
Hamiltonian $H_0$ with generalized complex eigenvalues ${\cal E}_m$.

Following Bohm and Gadella\cite{twentysix}, we will associate to the
resonant poles of the resolvent operator of the Schr\"odinger equation
of the unperturbed system, a rigged Hilbert space

\begin{equation}
\label{cuarentaicinco}\Phi _{+}\subset{\cal H}\subset \Phi ^*_{+},
\end{equation}

\noindent\
in this expression, $\Phi _{+}$ is the space of well-behaved functions
of the position coordinates ($\xi_i,\eta_j,r$) which are Hardy class 2
functions of the complex energy E from above, $\cal H$ is the Hilbert
space of square integrable functions and $\Phi^{*}_{+}$ is the space
of continuous antilinear functionals on $\Phi_{+}$. In this way, the
Gamow states of the unperturbed system are defined as continuous
antilinear functionals on $\Phi _{+}$, that is,

\begin{equation}
\label{cuarentaiseis}|\varphi _m(\xi_i,\eta_j,r)> \epsilon 
\Phi ^*_{+}.
\end{equation}

The adiabatically evolving Gamow functions $|\hat \varphi
_s(\xi_i,\eta_j,r;\vec \Re(t))>$ of the complete Hamiltonian $H$,
introduced in section 3, eq.(\ref{veintitres}), are linear
combinations of the Gamow functions $|\varphi _m(\xi_i,\eta_j,r)>$ of
$H_0$,

\begin{equation} 
\label{cuarentaisiete}|\hat \varphi _s(\xi_i,\eta_j,r;
[\vec \Re(t)])> = \sum _m|\varphi _m(\xi_i,\eta_j,r)>
{\bf K}_{ms}[\vec \Re(t)],
\end{equation}

\noindent\
where ${\bf K}[\vec \Re(t)]$ is the matrix which diagonalizes the
complex non-Hermitian matrix ${\bf H}[\vec \Re(t)]$.

Therefore, the adiabatically evolving Gamow functions are also
elements of $\Phi^*_{+}$,

\begin{equation}
\label{cuarentaiocho}|\hat \varphi _s(\xi_i,\eta_j,r;
[\vec \Re(t)])> \epsilon \Phi ^*_{+}.
\end{equation}

The dual of the Gamow eigenvector $|\hat \varphi _s[\vec \Re]>$ is the
eigenvector $<\hat \varphi _s[\vec \Re]|= \sum_m({\bf
  K}^{-1})_{sm}<\varphi_m(\xi_i,\eta_j)|$ defined in
eq.(\ref{veinticuatro}), corresponding to the same complex eigenvalue
${\cal E}_s[\vec \Re]$. Since, by assumption ${\bf H}[\vec \Re]$ has
no repeated eigenvalues for $\vec \Re \epsilon {\bf C}\subset M$, the
adiabatically evolving Gamow eigenvectors and their duals satisfy the
orthogonality relation

\begin{equation}
\label{cuarentainueve}<\hat \varphi_s[\vec \Re]|\hat \varphi_{s'}
[\vec \Re]> = 0,\hspace{1.5cm}\hat{\cal E}_s\neq \hat {\cal E}_s'
\end{equation}
inherited from the Gamow eigenfuctions $|\varphi_m(\xi_i,\eta_j,r)>$
of $H_0$, and they may be normalized to one

\begin{equation}
\label{cincuenta}<\hat \varphi _s[\vec \Re]|\hat \varphi_s
[\vec \Re] > = 1
\end{equation}

Having characterized the adiabatically evolving Gamow functions as
elements of a rigged Hilbert space we may turn to the question of the
geometric interpretation of the Berry phase factors of resonant
states.

\subsection{Adiabatic evolution and parallel translation}

Let us suppose that, as time varies, the self-adjoint Hamiltonian $
H[\vec \Re(t)]$ and its instantaneous energy eigenstates $\mid \hat
\varphi _s(\xi_i,\eta_j,r;\vec \Re(t))>$ make a cyclic excursion in a
closed circuit C in parameter space. We assumed that the time
dependence of the Hamiltonian justifies the adiabatic approximation.
We will further assume that, if $\vec \Re(t)$ is any point on C, the
complex matrix ${\bf H}[\vec \Re(t)]$ has no repeated eigenvalues in
an open neighbourhood of $\vec \Re \epsilon $ C, that is, the
Hamiltonian $H(t)$ has no repeated instantaneous complex eigenvalues
when $\vec \Re \epsilon $ C, and that both the generalized
eigenvectors $\mid \hat \varphi _s(\xi_i,\eta_j,r;\vec \Re(t))>$ and
the eigenvalues ${\cal E}_s[\vec \Re(t)]$ are smooth functions of
$\vec \Re \epsilon $C. Moreover, we will also assume that the
evolution of each state is such that there are no level crossings
along C.

To avoid a clumsy notation in the rest of this section we will not
write the position coordinates ($\xi_i,\eta_j,r$).

Let us call $F_s$ the complex line bundle defined by the Gamow
eigenfunction $\mid \hat \varphi _s[\vec \Re(t)]>$ over the parameter
space M of the adiabatically evolving system

\begin{equation}
\label{cincuentaiuno}F_s = \left\{(\vec \Re,\mid \hat \varphi _s 
[\vec \Re]>) |{ H} [\vec \Re]|\hat \varphi _s[\vec \Re]> = 
{\cal E}_s [\vec \Re ]|\hat \varphi _s[\vec \Re ]>, 
\vec \Re \epsilon  M \right\}
\end{equation}

Its fibre is a complex, one-dimensional linear space

\begin{equation}
\label{cincuentaidos}L^s_{\Re}: = \left\{|\hat \psi _s
[\vec \Re]>{\bf |} | \hat \psi _s[\vec \Re]> = e^{i\alpha }
|\hat \varphi _s[\vec \Re]>, |\hat \varphi _s[\vec \Re] > 
\epsilon  \Phi^*_{+},  \alpha [\vec \Re] \epsilon {\bf C^*},
\vec \Re \epsilon  M \right \}
\end{equation}

\noindent\
where $\Phi ^*_{+}$ is the space of antilinear functionals defined
over the space $\Phi _{+}$ of well-behaved functions of Hardy class 2
from above and $\alpha [\vec \Re]$ is a complex function of $\vec \Re
\epsilon M$.

Under the assumptions made above, the quantum number $s$, labelling
the eigenvalue $\hat {\cal E}_s[\vec \Re]$ and the eigenstate $|\hat
\varphi _s[\vec \Re]>$ of the instantaneous Hamiltonian, is an
adiabatic invariant. Hence, a system prepared in a state $|\hat \psi
_s [\vec \Re(t)]>$ such that

\begin{equation}
\label{cincuentaitres}|\hat \psi _s[\vec \Re(t_o)]> = C_s[
\vec \Re(t_o)]|\hat \varphi _s[\vec \Re(t_o)]>
\end{equation}
will evolve with ${H}[\vec \Re(t)]$ and be in a state

\begin{equation}
\label{cincuentaicuatro}|\hat \psi _s[\vec \Re(t)]> = 
C_s[\vec \Re(t)]|\hat \varphi _s[\vec \Re(t)]>
\end{equation}

\noindent
at $t$.

Now, consider the decomposition of a tangent vector in vertical and
horizontal parts

\begin{equation}
\label{cincuentaicinco}\frac{d}{dt}|\hat \psi _s(t)> = \left (
\frac{d}{dt}|\hat \psi _s(t)\right )_{\|} + \left (\frac{d}{dt}
\hat \psi _s(t)>\right )_{\bot}
\end{equation}

\noindent\
The horizontal part, orthogonal to the fibre is

\begin{equation}
\label{cincuentaiseis}\left(\frac{d}{dt}|\hat\psi _s(t)>
\right)_{\bot} = \sum _{m\neq s}<\hat \varphi_m[\vec \Re]
|\frac{d}{dt}|\hat \psi _s(t)>|\hat\varphi _m[\vec \Re]>
\end{equation}

\noindent\
and the vertical part, along the fibre is

\begin{equation}
\label{cincuentaisiete}\left (\frac{d}{dt}|\hat \psi _s(t)>
\right)_{\|} = <\hat \varphi _s[\vec \Re]|\frac{d}{dt}|
\hat \psi _s(t)>|\hat \varphi _s[\vec \Re]>
\end{equation}

\noindent\
If $|\hat \psi _s(t)>$ is an evolution state in an adiabatic change,

\begin{equation}
\label{cincuentaiocho}\left(\frac{d}{dt}|\hat\psi _s(t)>
\right)_{\|} = \left(\frac{dC_s[\vec \Re(t)]}{dt} + <\
hat \varphi _s[\vec \Re]|\frac{d}{dt}\hat \varphi _s[
\vec \Re(t)]>C_s[\vec \Re]\right)|\hat \varphi _s[\vec \Re]>
\end{equation}

The condition for parallel translation along the curve C is

\begin{equation}
\label{cincuentainueve}<\hat \varphi _s[\vec \Re]|\frac{d}
{dt}\hat \psi _s[\vec \Re]> = 0,\hspace{1.5cm} \vec \Re  
\epsilon  C
\end{equation}

\noindent\
which gives a one-form equation

\begin{equation}
\label{sesenta}dC_s[\vec \Re]  + <\hat \varphi _s[\vec \Re]
|d\hat \varphi _s[\vec \Re]>C_s[\vec \Re] = 0
\end{equation}

Its solution gives the Berry phase factor of the resonant state

\begin{equation}
\label{sesentaiuno}exp [{i \int _c i <\hat \varphi_s|d\
hat \varphi _s>}] = exp [{i\gamma _s[\vec \Re]}].
\end{equation}

In the case of a cyclic evolution, such that $\vec \Re(0) = \vec
\Re(T)$ and C is a closed curve, the complex phase $\gamma _s$ may be
expressed as

\begin{equation}
\label{sesentaidos}\gamma _s(C) = i\oint _c <\hat \varphi _s
|d\hat \varphi _s>.
\end{equation}

Now, it will be shown that the condition for the adiabatic evolution
of a resonant state is equivalent to the condition for parallet
translation along C.

The condition for the adiabatic time evolution of a resonant state,
equation (\ref{treintaitres}), may be written as

\begin{equation}
\label{sesentaitres}e^{\frac{-i}{\hbar}\int_{0}^{t} 
\hat {\cal E}_s(t')dt'}\left\{\frac{d}{dt}\left(
\hat {a}_s(t)e^{\frac{i}{\hbar}\int_{0}^{t} 
\hat{\cal E}_s(t')dt'}\right)  + <\hat \varphi_s|
\frac{d}{dt}\hat\varphi_s > \hat a_s(t) e^{\frac{i}{\hbar}
\int_{0}^{t} \hat{\cal E}_s(t')dt'}\right\} = 0.
\end{equation}

\noindent\
Now, making use of the normalization condition,
eq.(\ref{veinticinco}), this expressions may be written as

\begin{equation}
\label{sesentaicuatro}e^{-\frac{i}{\hbar}\int ^{t}_0{
\hat {\cal E}}_s(t')dt'}<\hat \varphi _s[\vec \Re(t)]|
\left\{|\hat \varphi _s[\vec \Re(t)]>\frac{dC_s(t)}{dt} + 
\frac{|d\varphi _s[\vec \Re(t)]>}{dt}C_s(t)\right\} = 0,
\end{equation}

\noindent\
where $C_s(t)$ is given by

\begin{equation}
\label{sesentaicinco}C_s(t) = e^{\frac{i}{\hbar}\int ^{t}_0
{\hat {\cal E}}_s(t')dt'}\hat {a}_s(t),
\end{equation}

\noindent\
multiplying both sides of (\ref{sesentaicuatro}) by the dynamic phase
factor, the condition for adiabatic evolution of a resonant state
takes the form

\begin{equation}
\label{sesentaiseis}<\hat \varphi _s|\frac{d}{dt}\hat \psi _s> = 
0,\hspace{1.5cm}\vec \Re  \epsilon  C
\end{equation}

\noindent\
which is the condition for parallel translation of the resonant state
$|\hat \psi _s[\vec \Re(t)]> = \hat C_s[\vec \Re(t)]|\hat \varphi
_s[\vec \Re(t)]>$ along the curve C.

It follows that, except for the effect of the dynamical phase factor,
the condition of adiabatic evolution of a resonant state is equivalent
to the condition of parallel translation on the complex line bundle
$F_s$.

The geometric phase factors $exp[i\gamma_s(c)]$, occurring in the
adiabatic evolution, are holonomy group elements of the complex line
bundle $F_s$ \cite{twentyfive}.

\section{Accidental degeneracy of resonances.}

As is apparent from (\ref{treintaiocho}), non-trivial phase factors of
the energy eigenvectors or eigenfunctions are related to the
occurrence of accidental degeneracies of the corresponding
eigenvalues. In the absence of symmetry, degeneracies are called
accidental for lack of an obvious reason to explain why two energy
eigenvalues ${\cal E}_s$ and ${\cal E}_m$ of $\bf H$ should coincide.
However, if the matrix $\bf H$ is embedded in a population of complex
non-Hermitian matrices $\left [ {\bf H}(X_1,X_2,...X_N)\right ]$
smoothly parametrized by $N$ external parameters $(X_1,X_2,....X_N)$,
degeneracy in the absence of symmetry is a geometric property of the
hypersurfaces representing the real or complex eigenvalues of $\bf H$
in a $(N+2)$-dimensional Euclidean space with Cartesian coordinates
$(X_1,X_2,....X_N,Re{\cal E}, Im{\cal E})$. The energy denominators in
(\ref{treintaiocho}) show that when the circuit $C$ lies close to a
subdomain $D'\subset D$, in parameter space, where the matrix $\bf H$
has an m-fold repeated eigenvalue, and the state $|\hat \varphi _s
\rangle$ is involved in this degeneracy, the Berry phase $\gamma
_s(c)$ is determined by the geometry of the energy hypersurfaces close
to the crossing of eigenvalues and the other (m-1) states involved in
the degeneracy.

It is in this connection of the Berry phase with the accidental
degeneracy of energy eigenvalues that the non-Hermiticity of the
matrix $\bf H$ plays an important role. In contrast with the case of
Hermitian matrices, square, complex, non-Hermitian matrices with
repeated eigenvalues cannot always be brought to diagonal form by a
similarity transformation. However, any n-dimensional, square complex
matrix $\bf H$ may always be brought to a Jordan canonical form $\bf
E$ by means of a similarity transformation.

If $\bf H$ has $\nu$ $(\nu \leq n)$ different eigenvalues, ${\cal
  E}_1, {\cal E}_2,....{\cal E}_{\nu}$, with multiplicities
${\mu}_i(E_i)$, the Jordan canonical form $\bf E$ is the direct sum of
$\nu$ square Jordan blocks ${\bf E}_i$. Each Jordan block ${\bf E}_i$
is the sum of a diagonal matrix ${\cal E}_i{\bf I}_{\mu_i\times
  \mu_i}$, and a nilpotent matrix $N_{\mu_i}$. Corresponding to each
Jordan block, there is a cycle of lenght $\mu_i$ of generalized
eigenvectors. When the lenght $\mu _i$ of the cycle is $\geq$2, the
codimension of the accidental degeneracy, the geometry of the energy
hypersurfaces at the crossing and the properties of the generalized
eigenvectors involved in the degeneracy of resonant states differ
substantially from those of bound states(\cite{thirteen}). Rather than
trying to develop a theory of the most general case, in the following
we will examine the simplest possible case, namely the accidental
degeneracy of two resonances and the topology of the energy surfaces
close to a crossing of two resonances in parameter space.

\section{Degeneracy of two resonances}

Let us consider a system with two resonant states strongly mixed by a
Hermitian interaction, all other bound or resonant eigenstates being
non-degenerate. We may suppose that we already know the correct
eigenvectors of $\bf H$ for all the real and complex eigenenergies
${\cal E}_s$, except for the two the crossing of which we want to
investigate. Using for this two states two vectors which are not
eigenvectors but which are orthogonal to each other and to all other
eigenvectors, we obtain a complete basis to represent $\bf H$. In this
basis, $\bf H$ will be diagonal except for the elements ${\bf H}_{12}$
and ${\bf H}_{21}$. The diagonal elements ${\bf H}_{11}$ and ${\bf
  H}_{22}$ will, in general, be non-vanishing and different from each
other. There is no loss of generality in this supposition, since any
complex matrix $\bf H$ may be brought to a Jordan canonical form by
means of a similarity transformation. When the eigenvalues are equal,
${\bf H}_{2\times 2}$ is either diagonal or equivalent to a Jordan
block of rank two. Hence, we need consider only the conditions for
degeneracy of the submatrix ${\bf H}_{2\times 2}$.

The matrix ${\bf H}_{2\times 2}$ may be written in terms of the Pauli
matrix valued vector $\vec \sigma =(\sigma _1,\sigma _2,\sigma _3)$
and the $2\times 2$ unit matrix as

\begin{equation}
\label{sesentaisiete}{\bf H}_{2\times 2}={\cal E}{\bf 1}+
{\left ({\vec R}-i\frac 12\vec \Gamma \right) \cdot \vec \sigma, } 
\end{equation}

\noindent
where $\vec R$ and $\vec \Gamma $ are real vectors with cartesian
components $\left( X_1,X_2,X_3\right) $and $\left( \Gamma _1,\Gamma
  _2,\Gamma _3\right) $. When the forces acting on the system are time
reversal invariant, $X_2$ and $\Gamma _2$ vanish.

In the absence of more specific information about the external parameters
$X$, we will parametrize ${\bf H}_{2\times 2}$ in terms of $\vec R$ and
$\vec \Gamma$. Then, 

\begin{equation}
\label{sesentaiocho}\vec \Re = \vec R - i \frac {1}{2}\vec \Gamma.
\end{equation}

{}From (\ref{sesentaisiete}), the eigenvalues of ${\bf H}_{2\times 2}$
are given by

\begin{equation}
\label{sesentainueve}{\bf E}_{1,2}={\cal E}\mp \epsilon . 
\end{equation}

\noindent
where

\begin{equation}
\label{setenta}\epsilon =\mp \sqrt{\left( \vec R-i\frac 12\vec \Gamma
\right) ^2}. 
\end{equation}

Then, ${\bf E}_1$ and ${\bf E}_2$ coincide when $\epsilon $ vanishes.
Since, real and imaginary parts of $\epsilon$ should vanish,the
condition for accidental degeneracy of the two interfering resonances
may be written as

\begin{equation}
\label{setentaiuno}R_d^2-\frac 14\Gamma _d^2=0, 
\end{equation}

\noindent
and

\begin{equation}
\label{setentaidos}\vec R_d\cdot \vec \Gamma _d=0. 
\end{equation}

These equations admit two kinds of solutions corresponding to ${\bf
  H}_{2\times 2}$ being or not being diagonal at the degeneracy:

i) When both $\vec R_d$ and $\vec \Gamma _{d}$ vanish, eqs.
(\ref{setentaiuno}) and (\ref{setentaidos}) define a point in
parameter space, ${\bf H}$ is diagonal at the degeneracy and the
submatrix ${\bf H}_{2\times 2}$ has two cycles of eigenvectors of
lenght one each. The vanishing of $\Gamma _d ^2$ implies that $Im
{\cal E}$ also vanish\cite{thirteen}, therefore the two complex
eigenvalues of $\bf H$ which become degenerate fuse into one real
positive energy eigenvalue embedded in the continuum. Since in this
case all the Cartesian components of $\vec R_d$ and $\vec \Gamma _d$
should vanish, the minimum number of linearly independent, real,
external parameters that should be varied to produce a degeneracy
(codimension) of two resonances to form a bound state embedded in the
continuum is four or six depending on the quantum system being or not
being time reversal invariant.

ii) In the second case, when the degeneracy conditions
(\ref{setentaiuno}) and (\ref{setentaidos}) are satisfied for
non-vanishing $\vec R_d$ and $\vec \Gamma _d$, these equations define
a circle in parameter space. In this case ${\bf H}_{2\times 2 }$ is
not diagonal at the degeneracy, it is equivalent to a Jordan block of
rank two and has one cycle of generalized eigenvectors of lenght two.
In this case, the two complex eigenvalues become one two-fold repeated
eigenvalue $\cal E$. Since, the two linearly independent conditions
(\ref{setentaiuno}) and (\ref{setentaidos}) should be satisfied for
non-vanishing values of ${\vec R}_d$ and ${\vec \Gamma}_d$, at least
two real, linearly independent parameters should be varied to produce
a rank two degeneracy of resonances. Hence, the codimension of a
degeneracy of resonances of second rank is two, independently of the
time reversal invariance character of the interactions.

\subsection{Computation of the geometric phase.}

Let us consider now the computation of the Berry phase in the simplest
situation when the degeneracy involves only two resonant states. The
matrix $\cal H$ that mixes the two interfering resonances off
degeneracy is

\begin{equation}
\label{setentaitres}
{\cal H}=\left( 
\begin{array}{cc}
Z-i\frac 12 \Gamma & X-iY \\ 
X+iY & -Z+i\frac 12 \Gamma 
\end{array}
\right) . 
\end{equation}
This matrix has two right and two left eigenvectors and may be
diagonalized by a similarity transformation

\begin{equation}
\label{setentaicuatro}
K^{-1}{\cal H}K=\left( 
\begin{array}{cc}
-\epsilon & 0 \\ 
0 & \epsilon 
\end{array}
\right) , 
\end{equation}
where

\begin{equation}
\label{setentaicinco}
K=\frac 1{\sqrt{2\epsilon }}\left( 
\begin{array}{cc}
\sqrt{\epsilon +\eta } & \sqrt{\epsilon -\eta } \\ 
\sqrt{\epsilon -\eta } \frac{ \xi ^{*}}{ \left| \xi 
\right| } & \sqrt{\epsilon +\eta }\frac{ \xi ^{*}}
{ \left| \xi \right| } 
\end{array}
\right) , 
\end{equation}
$\xi$ and $\eta$ are short hand for $X-iY$ and $Z-i\frac 12 \Gamma$
respectively.The matrices ${\bf K}^{-1}$ and $\nabla _R{\bf K}$ are
readily obtained from (\ref{setenta}) and (\ref{setentaicinco}). If we
assume that the interfering resonances are mixed by a Hermitian
interaction, $\vec \Gamma$ will be kept fixed. Then a straightforward
calculation gives

\begin{equation}
\label{setentaiseis}\gamma _1=-\frac 12\int_{{\bf c}}\frac 1
{\Gamma \epsilon \left( \epsilon +\eta \right) }\left( 
\vec \Gamma \times \vec R\right) \cdot d\vec R, 
\end{equation}
and

\begin{equation}
\label{setentaisiete}\gamma _2=-\frac 12\int_{{\bf c}}
\frac 1{\Gamma \epsilon \left( \epsilon -\eta \right) }
\left( \vec \Gamma \times \vec R\right) \cdot d\vec R. 
\end{equation}

These expressions are very similar to the well known results obtained
for the geometric phase of bound states\cite{two}. An obvious
difference is that the geometric phase of resonant states is complex
since $\epsilon $ and $\eta $ are complex functions of the parameters
$\vec R$ and $\vec \Gamma $. There is another important but less
apparent difference: In the case of an accidental degeneracy of
resonances $(\Gamma \neq 0),$ the denominator in the right hand side
of (\ref{setentaiseis}) and (\ref{setentaisiete}) vanishes on the
continuous line of singularities defined by eqs.(\ref{setentaiuno})
and (\ref{setentaidos}), which will be called the diabolical circle,
and not at one isolated point as is the case for bound states. It
follows that two kinds of non-trivial, topologically inequivalent
closed paths are possible. First, those paths which surround the
diabolical circle but are not linked to it. Second, the closed paths
which are linked to the diabolical circle. Paths of the first kind are
clearly analogous to the non-trivial paths that go around the
diabolical point of bound state degeneracies while paths of the second
kind have no analogue in accidental degeneracies of bound states.

\begin{figure}[htbp]
  \begin{center}
    \psfig{file=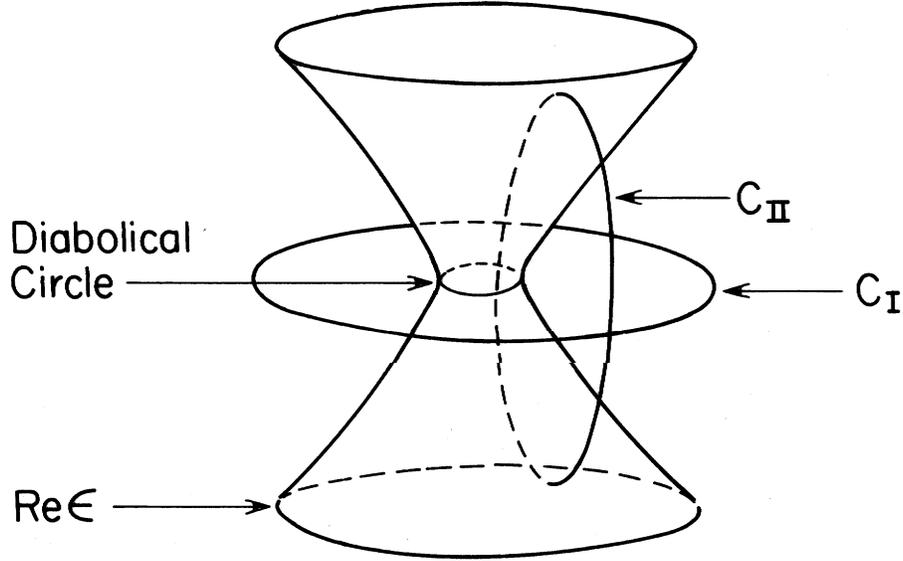}
    
    \caption{In the evaluation of the Berry phase of two interfering resonant
      states there are two kinds of non-trivial, topologically
      inequivalent closed paths in parameter space. First, those, like
      $C^{(I)}$, which go around the diabolical circle but are not
      linked to it. Second, those, like $C^{(II)}$, which turn around
      the diabolical circle and are linked to it.}
    \label{fig:1}
  \end{center}
\end{figure}

For paths of the first kind it is always possible to find a surface
$\Sigma $ which spans the closed path ${\bf C}$ and does not cross the
diabolical circle. Then, Stokes theorem applied to
(\ref{setentaiseis}) and (\ref{setentaisiete}) gives

\begin{equation}
\label{setentaiocho}\gamma _s=\frac{\left( -1\right) ^s}2
\int_{\Sigma_s}\int_{\partial \Sigma _s={\bf c}}\frac{
\left( \vec R-i\frac 12\vec \Gamma \right) \cdot d
\vec \Sigma}{\left[\left(\vec R-i\frac 12\vec \Gamma 
\right)^2\right ]^{\frac 32}}, 
\end{equation}
where $s=1,2.$ Since $\gamma _2$ changes into $\gamma _1$ when $\Sigma
_1$ and $\Sigma _2$ are exchanged and the sign of $d\vec \Sigma$ is
changed, the normals for $\Sigma _2$ and $\Sigma _1$ should be
oppositely oriented. If we say that $\vec \Gamma $ points upwards,
then $\Sigma _1$ is above and $\Sigma _2$ is below ${\bf C.}$

This is the same result as would have been obtained from the general
expression (\ref{treintaiocho}) and our parametrizacion of the
perturbation term in the Hamiltonian; no summation over intermediate
states occurs in (\ref{setentaiocho}) since in the simple case of only
two interfering resonant states, the summation in (\ref{treintaiocho})
has only one term.

Adding $\gamma _1$ and $\gamma _2$, we get

\begin{equation}
\label{setentainueve}\gamma _1+\gamma _2=-\frac 12\int_\Sigma 
\frac{\left( \vec R-i\frac 12\vec \Gamma \right) \cdot d\vec \Sigma}{
\left[\left(\vec R-i\frac 12\vec \Gamma\right)^2\right]^{\frac 32}}, 
\end{equation}
where $\Sigma $ is a sphere with the diabolical circle completely
contained in its interior. The integral is easily computed when $R>
\frac 12 \Gamma $, the result is

\begin{equation}
\label{ochenta}-\frac {1}{2\pi} (\gamma _1+\gamma _2)=1, 
\end{equation}

\noindent
in agreement with our identification of the integral in
(\ref{cuarentaicuatro}) as the first Chern number of the complex line
bundle defined by the eigenvector of the non-Hermitian matrix ${\bf
  H}_{2\times 2}$.

It is now easy to show that the resonance degeneracy produces a
continuous distribution of singularities on the diabolical circle. The
surface integral in (\ref{setentainueve}) may be written as a volume
integral by means of Gauss theorem. Then,

\begin{equation}
\label{ochentaiuno}\gamma _1+\gamma _2=-\frac 12\int \int 
\int_V\left(\nabla _R\cdot \frac{\left( \vec R-i\frac 12
\vec \Gamma \right) }{\left[\left( \vec R-i\frac 12\vec \Gamma 
\right) ^2\right] ^{\frac 32}}\right) dV, 
\end{equation}
where $V$ is the volume inside $\sum $ and bounded by it. The term in
round brackets under the integration sign vanishes when $\left (\vec
  R-i\frac 12 \Gamma \right )^2 \neq 0.$ Therefore, the non-vanishing
value of $\gamma _1+\gamma _2$ implies the occurrence of $\delta
$-function singularities of the integrand on those points where
$\left(\vec R-i\frac 12 \Gamma \right )^2 $ vanishes.

Hence,

\begin{equation}
\label{ochentaidos}\nabla _R\cdot \left[ \frac{\vec R-i
\frac 12\vec \Gamma }{\left[ \left( \vec R-i\frac 12
\vec \Gamma \right) ^2\right] ^{\frac 32}}\right] = -
\frac{\delta \left( R-\frac 12\Gamma \right) }{R^2}
\delta \left(\cos \theta \right), 
\end{equation}
the factor $R^{-2}$ multiplying the delta function is needed to
reproduce the value $2\pi $ of $\gamma _1+\gamma _2$.

Let us turn to the case of closed paths of the second kind, that is,
those paths which are linked to the diabolical circle. In this case
there is no surface $\Sigma$ which spans the closed path $c^{II}$
without crossing the diabolical circle. Therefore, it is not possible
to use Stokes theorem to convert the path integral into a surface
integral. However, we may still compute the geometric phase from the
path integral. To this end, we change from Cartesian coordinates
$(X,Y,Z)$ with $OZ$ parallel to $\vec \Gamma$, to spherical
coordinates in parameter space. Then, eqs. (\ref{setentaiseis}) and
(\ref{setentaisiete}) take the form

\begin{equation}
\label{ochentaitres}\gamma _{1,2}=-\frac 12\int_{{\bf c}}
d\varphi \mp \frac 12\int_{{\bf c}}\frac{\left( R\cos 
\theta -i\frac 12\Gamma \right) d\varphi }{\sqrt{R^2-
\frac 14\Gamma ^2-i\Gamma R\cos \theta }}, 
\end{equation}
the path $c$ is specified when $R$ and $\theta $ are given as
functions of $\varphi $.

{}From (\ref{ochentaitres}), it follows that, for closed paths which
are linked to the diabolical circle,

\begin{equation}
\label{ochentaicuatro}\gamma _1+\gamma _2=-\int_{c^{(II)}}d
\varphi =0,\qquad \qquad c^{(II)}{ of second kind} 
\end{equation}
since, in this case, the angle $\varphi$ starts out at some value
$\varphi _0$ and, as the system traces the path $c^{II}$, it
oscillates between a minimum and a maximum values and finally ends at
the same initial value $\varphi _0$. There is no analogue to this case
in the geometric phase of bound states.

\section{Results and conclusions}

The purpose of the foregoing has been to discuss the geometric phase
acquired by the resonant states when they are adiabatically
transported in parameter space around a degeneracy of resonances .

As in the case of bound states, the condition of adiabatic evolution
of resonant states may be given a geometric interpretation as parallel
translation in a complex line bundle $F_s$. The geometric phase
factors acquired by the resonant states when adiabatically transported
in a closed circuit in parameters space are holonomy group elements of
the complex line bundle $F_s$.

In the case of two resonant states mixed by a Hermitian interaction we
find two kinds of accidental degeneracies which may be characterized
by the number and length of the cycles of instantaneous energy
eigenfunctions at the degeneracy. In the first case there are two
linearly independent eigenfunctions belonging to the same real
positive repeated energy eigenvalue, that is, two cycles of lenght
one. In the second case there is only one resonant eigenstate and one
generalized resonant eigenstate belonging to the same degenerate
(repeated) complex energy eigenvalue, $i.e$. one cycle of lenght two.
At degeneracy, the Hamiltonian matrix has one Jordan block of second
rank.

In the generic case of a non-time reversal invariant system, when the
degeneracy is of the second rank, the topology of the energy surfaces
is different from that at a crossing of bound states. The energy
surfaces of the two resonant states that become degenerate touch each
other at all points in a circle. Close to the crossing, the energy
hypersurface has two pieces lying in orthogonal subspaces in parameter
space. The surface representing the real part of the energy has the
shape of a hyperbolic cone of circular cross section, or an open
sandglass, with its waist at the diabolical circle. The surface of the
imaginary part of the energy is a sphere with the equator at the
diabolical circle. The two surfaces touch each other at all points on
the diabolical circle\cite{thirteen}.

\begin{figure}[htb]
  \begin{center}
   \psfig{file=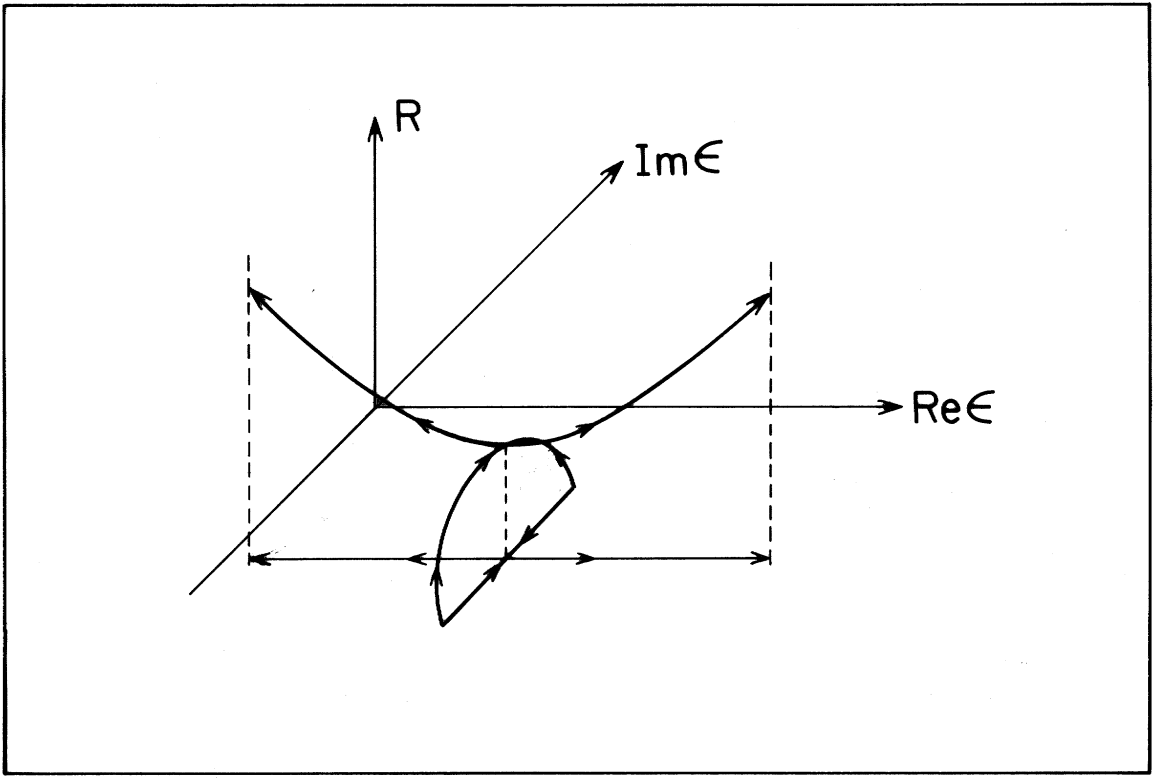}
    
    \caption{Two interfering resonances which initially have equal level
      energies but different half widths are mixed by the Hermitian
      interaction $\vec R\cdot \vec \sigma$. As $R$ increases from 0
      to $\frac{1}{2}\Gamma$, the points representing ${\cal E}_1$ and
      ${\cal E}_2$ approach each other along a meridian circle on the
      sphere representing $Im \epsilon $ in parameter space, until
      they meet at a point on the equator, which is the diabolical
      circle. At this point the two resonances become degenerate and
      the mixing matrix ${\bf H}_d$ is equivalent to a Jordan block of
      rank two. When $R$ becomes larger than $\frac{1}{2}\Gamma$ the
      points representing ${\cal E}_1$ and ${\cal E}_2$ move away from
      each other on the hyperbolic cone representing $Re \epsilon$.}
    \label{fig:2}
  \end{center}
\end{figure}

In the case of two interfering resonant states, the geometric phase
acquired by the resonant states when transported around the diabolical
circle in a closed path which is not linked to it, may be written as
the sum of two terms.

\begin{equation}
\gamma _{1,2}^{res}\left( {\bf c}^I\right) =\gamma
_{1,2}^{bound}\left( {\bf c}^I\right) \pm \Delta \gamma 
\left( {\bf c}^I\right). 
\end{equation}

The first term, $\gamma _{1,2}^{bound}\left( {\bf c}^I\right) $, is
the real geometric phase which a negative energy eigenstate would have
acquired when transported around a diabolical point in a closed path
in the same parameter space. The second term is complex, it gives rise
to a change of the phase and a dilation of the resonant state
eigenfunction. Its imaginary part may be positive or negative, in
consequence, it may produce an amplification or a damping of the wave
function which may compensate or reinforce the attenuation due to the
imaginary part of the dynamical phase factor.

When the resonant states are transported in a closed path ${\bf
  c}^{II}$ which does not go around the diabolical circle but is
linked to it, the geometric phase they acquire is

\begin{equation}
\gamma _{1,2}^{res}\left( {\bf c}^{II}\right) =\pm
\Delta \gamma \left( {\bf c}^{II}\right). 
\end{equation}

Since it is not possible to find a continuous surface $\sum $ which
spans the closed path ${\bf c}^{II}$ without crossing the diabolical
circle, we can not make use of the theorem of Stokes to convert the
path integral into a surface integral. However, it may readily be
computed as a path integral from the expression

\begin{equation}
\Delta \gamma \left( {\bf c}^{II}\right) =\int_{{\bf c}
^{II}}\frac{\left( Z-i\frac 12\Gamma \right) \left( 
\vec \Gamma \times \vec R^{\prime }\right) \cdot d
\vec R^{\prime }}{\Gamma \sqrt{\left( \vec R-i\frac 12
\vec \Gamma \right) ^2}\left( X^2+Y^2\right) }, 
\end{equation}

\noindent\
which is obtained from (\ref{ochentaitres}) and
(\ref{ochentaicuatro}). As in the previous case, $\Delta \gamma \left(
  {\bf c}^{II}\right) $ is complex and produces changes of phase and
dilations in the resonant state wave function. This case has no
analogue in bound states.

The sum of the geometric phases acquired by two interfering resonant
states which are transported around a degeneracy in a closed path of
the first kind in parameter space is a topological invariant, namely
the first Chern number\cite{twentyfive}. Its value is the ``magnetic
charge'' on the diabolical circle. For paths of the second kind the
sum of the geometric phases vanishes.

\end{document}